\documentclass{INTERSPEECH2023}
\usepackage{multirow}

\interspeechcameraready

\title{\vspace{-0.2cm} MERLIon CCS Challenge: A English-Mandarin code-switching child-directed speech corpus for language identification and diarization}

\name{Victoria Y. H. Chua$^{1*}$, Hexin Liu$^{2*}$, Leibny Paola Garcia Perera$^3$, Fei Ting Woon$^1$, Jinyi Wong$^1$, Xiangyu Zhang$^3$, Sanjeev Khudanpur $^3$, Andy W.~H.~Khong$^2$, Justin Dauwels$^4$, Suzy J.~Styles$^1$}

\address{\vspace{-0.2cm}
  $^1$Psychology, School of Social Sciences, Nanyang Technological University, Singapore\\
  $^2$School of Electrical and Electronic Engineering, Nanyang Technological University, Singapore\\
  $^3$CLSP and HLT-COE, Johns Hopkins University, USA\\
  $^4$Department of Microelectronics, Delft University of Technology, Netherlands
}
\email{victoriachua@ntu.edu.sg, hexin.liu@ntu.edu.sg, lgarci27@jhu.edu}

\begin{document}

\maketitle

\begin{abstract}
 To enhance the reliability and robustness of language identification~(LID) and language diarization~(LD) systems for heterogeneous populations and scenarios, there is a need for speech processing models to be trained on datasets that feature diverse language registers and speech patterns. We present the MERLIon CCS challenge, featuring a first-of-its-kind Zoom video call dataset of parent-child shared book reading, of over 30 hours with over 300 recordings, annotated by multilingual transcribers using a high-fidelity linguistic transcription protocol. The audio corpus features spontaneous and in-the-wild English-Mandarin code-switching, child-directed speech in non-standard accents with diverse language-mixing patterns recorded in a variety of home environments. 
 This report describes the corpus, as well as LID and LD results for our baseline and several systems submitted to the MERLIon CCS challenge using the corpus. 
\end{abstract}

\vspace{-0.2cm}
\section{Introduction}
In recent years, language identification and diarization have progressed significantly, resulting in models that achieve low error rates across different languages \cite{lre_eval, efficient_lid}. However, most speech processing technologies have been developed for performance on a narrow range of audio data – adult voices, speaking or reading a single language in a particular variety, recorded in a controlled acoustic environment \cite{librispeech, ami}. However, spontaneous speech in-the-wild is rarely so controlled (see CHiME data \cite{chime5}). In addition, speaking in more than one language, switching rapidly between languages, using a variety of English other than American or British English, and adopting different speech registers when speaking to different parties, are common for most speakers in the globalized world. Recent work has found that language identification systems struggle to cope with different varieties of English \cite{kukk22_interspeech, chan22b_interspeech, zhang22n_interspeech}, as well as spontaneous in-the-wild speech containing more than one language and extremely short language spans \cite{liu21_interspeech, liu22_interspeech, liu22c_odyssey, bi_encoder}. Hence, there is a need to develop models that can handle spontaneous code-switched speech across different populations of speakers.

While speech technologies have been mainly trained on adult voices, effective technology requires robustness across speech registers (e.g., when the same speaker adopts a different style and pitch for a different audience). For instance, child-directed speech by adults comprises a range of acoustic features that differentiate it from adult-directed speech, such as overall higher fundamental frequency, expanded pitch range, hyperarticulation of vowel formants and tones, slower speech rate and lengthened speech segments. These features present critical challenges for automatic language identification and language diarization. Speech diarization systems when employed for parent-child conversations often underestimate the number of child vocalizations, child-adult, and adult-adult conversation turns ~\cite{cristia18_interspeech,cristia2021thorough}. Moreover, many features of child-directed speech also occur in other low-intelligibility speech contexts including speech produced for foreigners \cite{UTHER20072}, for artificial speech processing systems \cite{robot} and when audio clarity is reduced ~\cite{hazan2015, Burnham2002}. Hence reducing errors on speech containing these features will have impacts beyond child-directed speech alone. Notwithstanding the above, with the increase in the ubiquity of videocalls, conversations recorded on videocall platforms across a variety of home environments (user devices, connection speeds, and noise types) are understudied in the field of speech processing. 

To enhance the inclusiveness, reliability, and robustness of language identification and diarization systems for heterogeneous populations and scenarios, there is a need for systems to be trained on datasets that feature different registers and speech patterns (e.g., US/UK English and Mainland Chinese), and spontaneous codeswitching. To that end, we present a unique audio dataset of Zoom videocalls featuring a South-east Asian variety of English-Mandarin code-switched child-directed spontaneous speech with diverse language-mixing patterns across the speakers recorded in a variety of home environments, curated for the Multilingual Everyday Recordings - Language Identification on Code-Switched Child-Directed Speech (MERLIon CCS) challenge\footnote{https://github.com/MERLIon-Challenge/merlion-ccs-2023}. In this report, we describe the MERLIon CCS challenge, with a detailed description of the dataset, baseline experiments and the results of systems submitted to the challenge.

\vspace{-0.2cm}
\section{Dataset Description}
The MERLIon CCS Challenge dataset \cite{merliondata} results from the collaboration between researchers investigating children's language development in multilingual scenarios and speech engineers working on spontaneous and code-switched speech. 
\vspace{-0.2cm}
\subsection{Data Collection}
The audio data collection was part of the Talk Together Study, a large developmental study conducted in Singapore \cite{woon2021creating}. The study is approved by the Nanyang Technological Institutional IRB Board (IRB-2018-10-001). In the study, parents were required to narrate an onscreen wordless picture book \cite{styles2020little} to their children over the Zoom video-conferencing software. All participating parents gave consent for the video calls (including their child's voice) to be recorded and audio to be released. Each recorded session lasted between 3 and 35 minutes, of which 2 to 25 minutes have been manually annotated by a team of multilingual transcribers \cite{bela}. The MERLIon CCS Challenge dataset includes only audio recordings. 

The entire dataset (including development and evaluation datasets) contains 305 Zoom audio recordings of 112 parent-child pairs \cite{merliondata}. 
103 of the parent-child pairs were recorded at least twice on separate occasions, with a maximum of three recordings for each pair. The dataset includes over 25 hours of English child-directed speech and over 5 hours of Mandarin Chinese child-directed speech. Around 10\% of the speech in the dataset is from children. Almost all adult voices in the dataset are female, speaking to children under the age of 5.  

\begin{figure}[t]
  \centering
  \includegraphics[width=\linewidth]{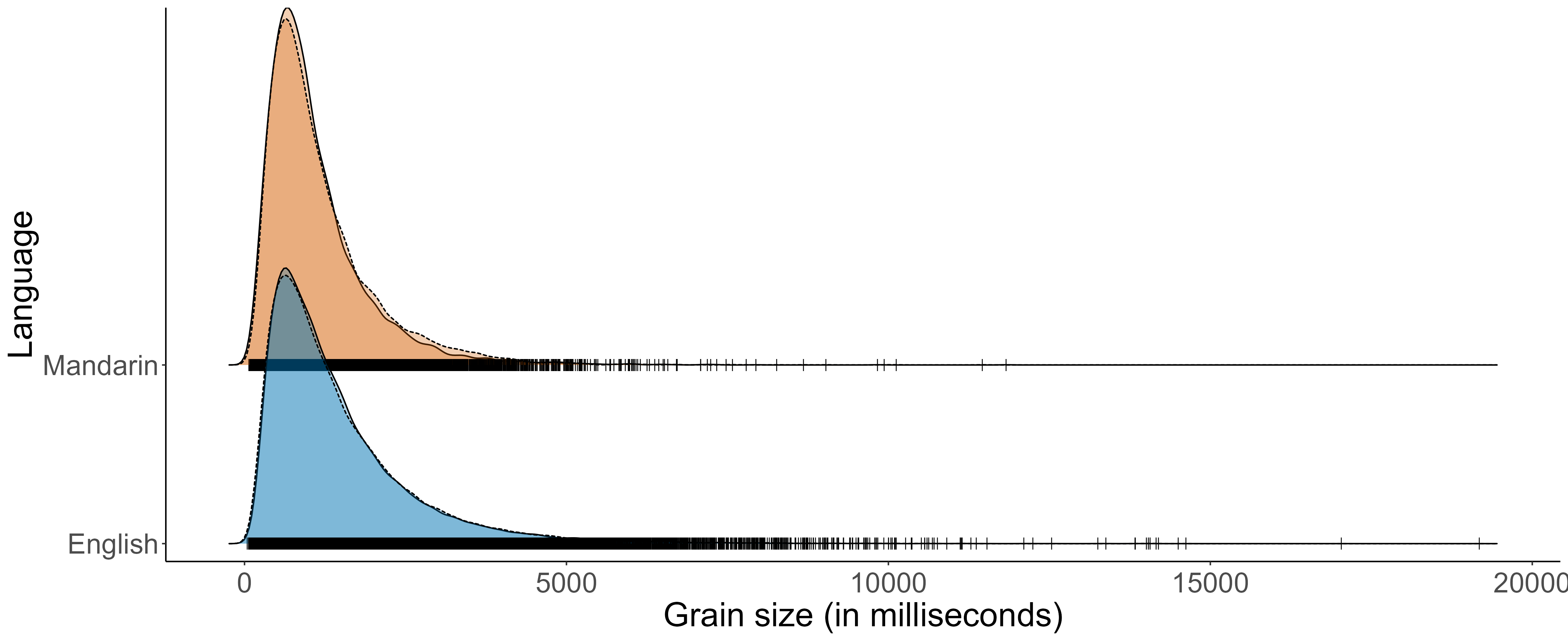}
  \includegraphics[width=\linewidth]{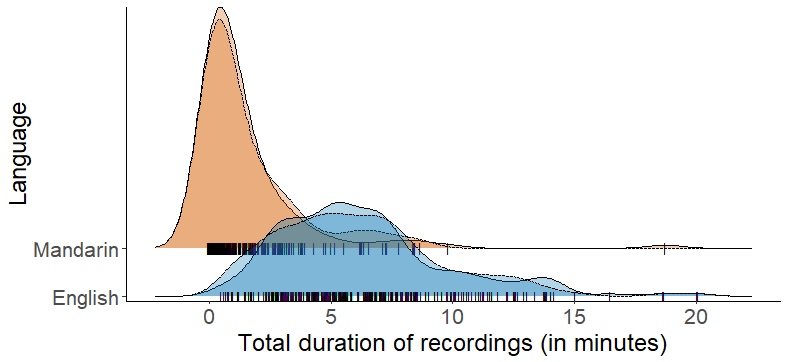}
  \caption{Density plots comparing English and Mandarin audio recordings in Development (Solid line) and Evaluation (Dashed line) sets, with strip plots indicating distribution. Top. Grain size over Development (N = 9983 Mandarin grains; 40287 English grains) and Evaluation (N = 9766 Mandarin grains; 39473 English grains) sets. Bottom. Total duration of each language in recordings over Development ($N=151$) and Evaluation ($N=154$) sets.}
  \label{fig:r_plot}
\vspace{-0.4cm}
\end{figure}


Zoom calls were conducted in the homes of participating families on a variety of internet-enabled personal electronic devices including laptops, tablets, and mobile phones.\footnote{Although some metadata is available for some recordings, the Zoom platform does not contain verifiable data about user services.} Environmental background noise and voices of other family members are also presented in the recordings.  
\vspace{-0.2cm}
\subsection{Languages and Accents}
Adults in our dataset use the Singaporean variety of English \cite{deterding2007singapore}, which features different pronunciations from Standard US, Standard UK, and other well-documented varieties of English, and the Singaporean variety of Mandarin Chinese, which features different pronunciations from the standard variety of Mandarin (Putonghua), and other well-documented Chinese varieties~\cite{lee2010tonal}. The Singaporean varieties also feature some unique vocabulary and grammar \cite{deterding2007singapore}. 
The dataset includes frequent \emph{code-switching} within and between utterances. 

Only 61 recordings (20\%) feature one language throughout. For parents who used both languages, the proportion of Mandarin spoken overall ranged from 0.85\% to 80.7\%. The utterances are short (mean of 1.4 seconds for English and 1.1 seconds for Mandarin). A breakdown of the composition of files in the Development and Evaluation sets can be found in Table~\ref{tab:2}.
\vspace{-0.2cm}
\subsection{Human Annotation}
As part of the main study, each audio recording was manually annotated by transcribers in ELAN, using the in-house transcription protocol \cite{bela}. As part of this protocol, transcribers were instructed to segment and annotate all utterances (i.e., any sound/voice produced by the speaker via their vocal apparatus). The start and end of an utterance are defined by intonation patterns and pauses \cite{bela}. In the transcription protocol, a subdivision of an utterance, due to code-switching to a different language, is known as a ``grain". Onsets and offsets of different languages are marked. In addition, each utterance or subdivision is labelled with boundaries for non-linguistic communicative acts including vocal sounds (e.g., humming) and non-vocal sounds (e.g., clapping). 

Transcribers were given instructions to place the start-stop boundaries carefully, taking note to include sounds at the edge of words (such as fricatives like /s/ at the end). In the event of a language change, they were asked to include all word boundaries in each language. When overlap between speakers occurs, transcribers are instructed to identify the start and end of each speaker turn to the best of their abilities. To minimize transcription errors, each file in the dataset has been crosschecked by at least one senior member of the transcription team. 

\begin{table}[t]
  \caption{Training datasets for the closed track}
  \label{tab:1}
  \centering
  \setlength{\tabcolsep}{1.7mm}{
  \begin{tabular}{c|c|c}
    \toprule
    \multicolumn{1}{c|}{\textbf{Corpus}} &  \multicolumn{1}{c|}{\textbf{Hours}}& \multicolumn{1}{c}{\textbf{Language}}\\ 
    \midrule
    LibriSpeech   & 100 & English      \\
    AISHELL      & 200 & Mandarin     \\
    NSC           & 101  & English     \\
    SEAME         & 192  & English-Mandarin Code-Switching  \\
    \bottomrule
  \end{tabular}}
\vspace{-0.4cm}
\end{table}
\vspace{-0.2cm}
\section {MERLIon CCS Challenge}
The Evaluation dataset is selected to be a representative subset of the Development dataset, where features such as ratio of English to Mandarin per recording are well-matched as shown in Figure~\ref{fig:r_plot} and Table~\ref{tab:2}. To reduce overfitting to individual parent-child pairs, the pairs in the Evaluation dataset do not appear in the Development dataset. Voices of some research assistants who were initiating the videocall repeat in both Development (4.1\%) and Evaluation (4.5\%) datasets.

\vspace{-0.2cm}
\subsection{Open and Closed Datasets for Training}
For the MERLIon CCS Challenge, we created two tracks which differ in terms of the volume of data systems can be trained on. We refer to these tracks as open versus closed respectively. 

In the closed track, to facilitate replicable research outcomes and for systems to be broadly comparable across research communities, we specified only training on pre-selected partitions of three open-access monolingual speech corpora and one LDC corpus (Table 1)\footnote{Access to pre-selected partitions at https://sites.google.com/view/merlion-ccs-challenge/datasets.}. The motivation behind using a limited amount of monolingual training data is to determine the lower limit of error rates when dealing with complex code-switching speech with a fixed set of monolingual data resources for training. In many contexts, monolingual speech is better represented in existing speech corpora and may be less challenging to collect and annotate. In the open track, all systems were allowed an additional maximum 100 hours of any publicly available or proprietary data. Pretrained models that are publicly available were also allowed in the open track.  

\begin{table}[t]
  \caption{Data distribution in Development and Evaluation sets}
  \label{tab:2}
  \centering
  \footnotesize
  \setlength{\tabcolsep}{1.7mm}{
  \begin{tabular}{l|c|c}
    \toprule
    \multicolumn{1}{l|}{} &  \multicolumn{1}{c|}{\textbf{Dev}}& \multicolumn{1}{c}{\textbf{Eval}}\\ 
    \midrule
    Total duration (hh:mm:ss)    & 28:36:28 & 28:47:14       \\
    Recordings N  & 151 & 154     \\
    Parent-child pairs N & 56 	&56   \\
    Recordings with one language N & 28 	&33   \\
    Total English speech (hh:mm:ss) 	&16:09:27 	&15:57:02 \\
    Total Mandarin speech (hh:mm:ss) 	&03:04:05 	&03:11:04 \\
    Total other language speech (hh:mm:ss)	&00:00:14 	&00:00:19 \\
    English segments N 	&40287 	&39473 \\
    Mandarin segments N &9983 	&9766 \\
    Median length of English segments~(ms) 	&1125 	&1120 \\
    Median length of Mandarin segments~(ms) &900 	&930 \\
    Mean length of English segments~(ms)	&1443.84 &1454.73\\
    Mean length of Mandarin segments~(ms) 	&1106.41 &1173.87\\
    Mean Proportion of English / Total  & 0.85 &0.83\\
    Median Proportion of English / Total  & 0.95 &0.94\\
    \bottomrule
  \end{tabular}}
\vspace{-0.4cm}
\end{table}

\vspace{-0.2cm}
\subsection{Task 1: Language Identification}
For language identification, the goal is to detect and label the language spoken automatically in a given audio segment marked by ground-truth timestamps of an audio recording. During development, systems were provided with audio recordings where ground-truth language labels have been annotated with timestamps. Each audio segment had a unique language label of either English or Mandarin spoken by the parent, child or research assistant. During evaluation, audio recordings were provided with timestamps of audio segments, but the language labels of these audio segments were not provided.

As the language identification task is binary, the primary and secondary evaluation metrics were system-level equal error rate~(EER) and balanced accuracy, respectively. Only English and Mandarin segments within each recording were scored during evaluation. Overlapping language segments with different language labels were not scored. For instance, for an English speech segment overlapping with another Mandarin speech segment, both speech segments were excluded from scoring. By contrast, English and Mandarin segments that overlap with non-speech segments were scored. In the event of overlapping speech, if the languages of all speakers are the same, the audio segment was scored.
\vspace{-0.2cm}
\subsection{Task 2: Language Diarization}
The goal of language diarization is to detect spans of each language spoken in each audio recording automatically with no pre-established timestamps for speech or language. During development, systems were provided with audio recordings with ground-truth timestamps and language labels. During evaluation, only the audio recordings were provided. To ensure a fair evaluation of conversational code-switching speech, language overlaps of different language labels were evaluated. In the event of overlapping speech, where languages were the same, both speech segments were evaluated. When a speech segment occurs with another speaker’s non-speech vocalizations, only the region which contains the evaluated language segment (i.e., English or Mandarin) was considered for evaluation. For operational reasons in the primary study, for each recording, there were some regions where no annotations have been performed, i.e., there may be speech unannotated for timestamps and language labels. Timestamps of these regions in the Development set were provided as well. These regions were excluded from the test audio during evaluation.


The primary and secondary evaluation metrics were system-level language diarization error rate (LDER) and individual language error rates (LER), respectively. The LDER is based on the speaker diarization error rate used in speech diarization system evaluations \cite{dihard}. The LDER is computed as the sum of:  
\begin{itemize}
    \item Language error – percentage of scored time for which the wrong language tag is assigned to a speech region.
    \item False alarm speech – percentage of scored time for which a nonspeech region is incorrectly marked as English or Mandarin speech.
    \item Missed speech – percentage of scored time for which English or Mandarin speech is incorrectly marked as nonspeech.   
\end{itemize}

For each system, the individual LERs for English and Mandarin are computed as the percentage of scored time for which a speech region containing the target language is incorrectly marked as non-speech or another language. 

\vspace{-0.2cm}
\subsection{MERLIon CCS Baseline System}
The baseline system
is an end-to-end conformer model. The conformer model has shown higher performance in both language identification and speech recognition tasks \cite{liu22c_odyssey, wang22b_odyssey}. The baseline model consists of four conformer encoder layers followed by a statistics pooling layer and three linear layers with ReLU activation in the first two linear layers. All self-attention encoder layers have eight attention heads with input and output dimensions being 512, and the inner layer of the position-wise feed-forward network is of dimensionality 2048. The 39-dimensional mel-frequency cepstral coefficients~(MFCCs) features comprising 13-dim MFCCs and their first- and second-order deviations are extracted for each speech signal before being fed into the conformer encoder layers. The statistics pooling layer then generates a 1024-dimensional output which is finally projected by three linear layers, comprising 1024, 512, and 2 output nodes, to the number of target languages. 

The baseline system is trained on the pre-selected partitions of three monolingual speech corpora (Table 1). The speech signals in these datasets were segmented into a maximum of 3~s prior to the feature extraction stage. The model was trained for five epochs with batch size 32 and updated with a learning rate that warms up from 0 to $10^{-4}$ in 5000 steps followed by the cosine annealing decay. For language diarization, an energy-based voice activity detection is performed to identify the silent parts. Each speech signal is partitioned into speech clips after removing silences before performing language identification on these clips. We assume no code-switch exists in each speech clip. The performance of our baseline system is summarized in Table 3.

\begin{table}[t]
  \caption{Baseline system results}
  \label{tab:3}
  \centering
  \begin{tabular}{c|c|c}
    \toprule
    \multicolumn{1}{c|}{\textbf{Metric}} &  \multicolumn{1}{c|}{\textbf{Dev}}& \multicolumn{1}{c}{\textbf{Eval}}\\ 
    \midrule
    Equal Error Rate~(EER)   & 22.1\% & 21.7\%      \\
    Balanced Accuracy~(BAC)  & 50.3\% & 50.9\%     \\
    Language Diarzation Error Rate & 86.6\%  & 84.0\%     \\
    English LDER       & 83.9\%  & 80.5\% \\
    Mandarin LDER      & 99.8\%  & 101.2\% \\
    \bottomrule
  \end{tabular}
\vspace{-0.4cm}
\end{table}

\vspace{-0.2cm}
\section{Results and Discussion}
\begin{figure}[t]
  \centering
  \includegraphics[width=\linewidth]{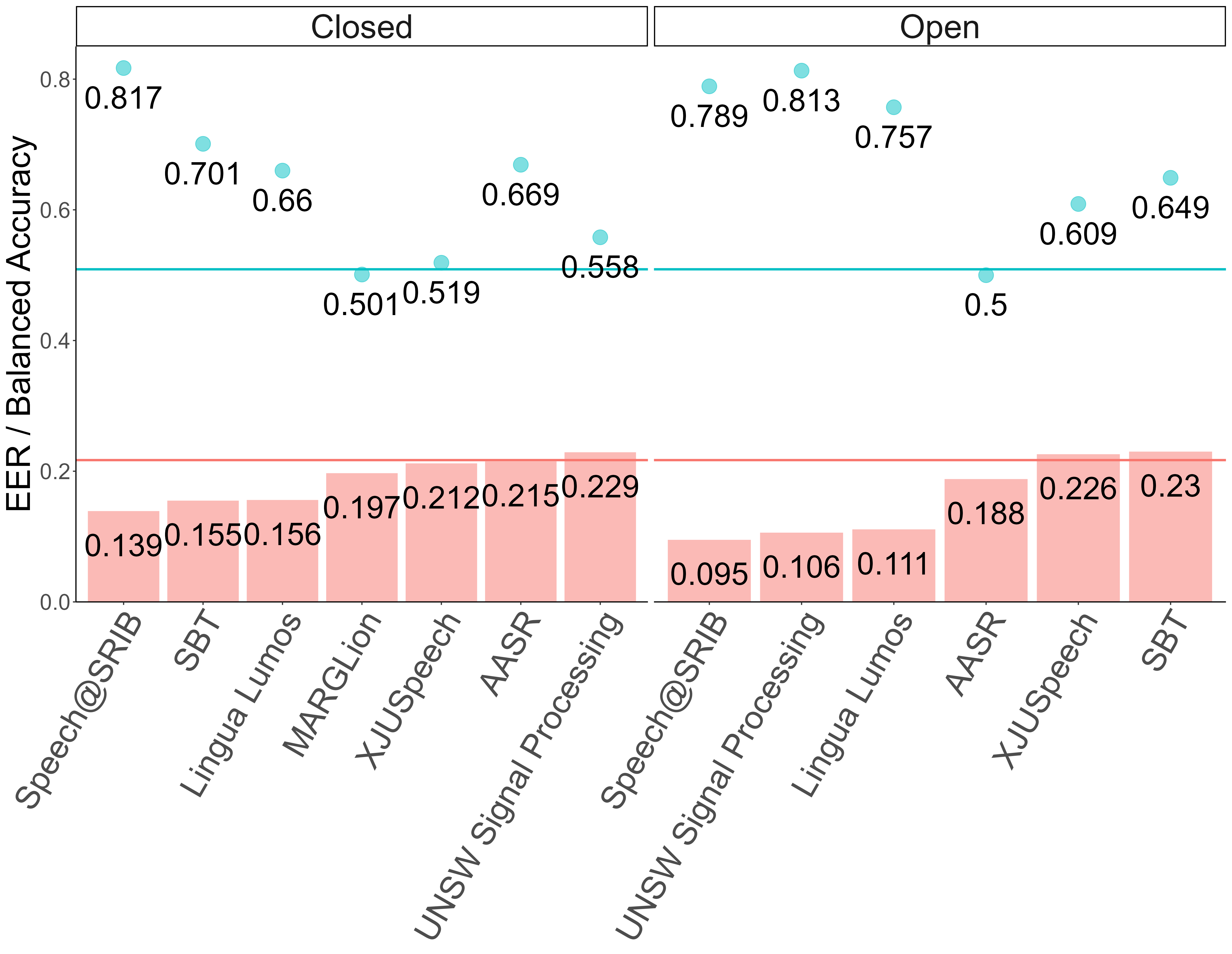}
  \caption{Equal error rates (bars) and balanced accuracy (dots) of language identification systems submitted to closed and open tracks compared to baseline EER and BAC (lines).}
  \label{fig:results}
\vspace{-0.6cm}
\end{figure}

The MERLIon CCS Challenge was outlined in the evaluation plan \cite{evalplan} and all submitted system outputs were evaluated via CodaLab \cite{codalab_competitions}. We compare our baseline system language identification results to 7 systems submitted to the closed track and 6 submitted to the open track (Figure~\ref{fig:results}). No systems were submitted to the language diarization task. All submitted models made improvements over the baseline system in terms of EER, BAC or both. The three best performing models in the Open track outperformed all models in the Closed track in terms of EER, of which two of these were adaptations of systems submitted to the closed track. 

The first place winner in the Open track achieved an EER of 9.50\% and a BAC of 78.9\%. Submitted by Speech@SRIB (Samsung R\&D Institute India-Bangalore), the system had an ensemble learning method combining a multilingual ASR network with the pre-trained Whisper-large-v2 \cite{whisper}.
The multilingual ASR network is a transformer-based model trained on the monolingual datasets in Table 1 and the Development set with RNN-T loss. This model is adapted to generate the language tag before generating the linguistic tokens. The ASR network is then trained on a subset of the Development set with weighted cross-entropy loss applied to the output layer. The final system combines language posteriors from the multilingual ASR model and Whisper model. In the closed track, the fine-tuned multilingual ASR system alone ranked first, with an EER of 13.9\% and 81.7\% BAC.  

The University of New South Wales (UNSW) Signal Processing team submitted the second placed system in the Open track. They fine-tuned the wav2vec2-large-robust 
model \cite{wav2vec} with phonological attributes from the Librispeech partition (Table~\ref{tab:1}) extracted by the CMU dictionary. To handle short speech segments, truncated utterances of 3 seconds from the Development set were used for downstream fine-tuning in the language identification phase. The system achieved an EER of 10.6\% and 81.3\% BAC. 

The third place system in the Open track was submitted by Lingua Lumos team from RingCentral and utilized the pre-trained titanet-1 model~\cite{titanet} in the Nvidia NeMo toolkit~\cite{nemo} and handles the problem of varying speech segment lengths and accented speech with an additional curated dataset. Six models based on the TitaNet architecture were trained on varying segment lengths from the closed training dataset with combinations of loss functions to increase model variability, forming an ensemble system. Then, speech segments from Singaporean English and Mandarin YouTube videos and speech of less than 6 seconds from the Mozilla CommonVoice dataset \cite{ardila-etal-2020-common} were selected. Speech segments misclassified by this ensemble system formed a curated dataset of 44 hours. Finally, fine-tuning of 7 different titanet-1 models was performed using the LDC SEAME corpus, the Development set, and the curated dataset of misidentified segments. The final ensemble 7-model, achieves an EER of 11.1\% and 76.0\% BAC. In the Closed track, the ensemble 6-model system alone ranked third, with an EER of 15.6\% and 66.0\% BAC. 

In the Closed track, in addition to the first and third placing systems already described, the second place system pre-trained a bilingual Conformer ASR model on monolingual datasets in Table 1, then fine-tuned on the Development set and segments derived from code-switched utterances in LDC SEAME corpus. This system, submitted by SBT from John Hopkins University, achieved an EER of 15.5\% and 70.1\% BAC. 

The top three systems across both tracks incorporated pre-trained models. The challenge saw improvements to developing language models that handle complex code-switched accented speech. The best performing system achieved an error rate below 10\% and improvements of 12.2\% over baseline EER and 30.4\% over baseline BAC. In line with the rules of the Challenge, full model descriptions of all submitted systems can be found at the MERLIon CCS Challenge GitHub site.
\vspace{-0.2cm}
\section{Conclusion}
Tests performed on the MERLIon CCS evaluation dataset suggest that large pre-trained models can be useful for language identification tasks that feature code-switching speech in different accents and registers when used in combination with models that are trained or curated via a small amount of data that match the speech characteristics of the targeted dataset. Our results highlight the possibility that at least 30 hours of fine-tuning data is beneficial for large pre-trained models to adapt. 
We further investigate the impact of significantly short speech segments (e.g., less than 3 seconds), presence of local vernacular (e.g., discourse markers like ``lah'' \cite{discoursemarkers}), and rate of code-switching on language identification performance in a separate paper \cite{merlionerrors}.

It should be acknowledged that language diarization remains a complex task, particularly for recordings across diverse recording environments and noise sources. Such requirements give rise to a robust speech detection system that can accurately generate speech onset and offsets before language identification can occur. This also accounts for why we received sign-ups for our language diarization task but no submissions. As feedback from participants suggests, significant difficulties in identifying the on and offset timings of the speech markers affected language diarization performance. For future research, there is a need to benchmark existing speech detection and diarization tools on our datasets so as to ascertain its errors and impact on subsequent language diarization performance.
\vspace{-0.2cm}
\section{Acknowledgements}
This work was partially supported by the National Research Foundation, Singapore, under the Science of Learning programme (NRF2016-SOL002-011), the Centre for Research and Development in Learning (CRADLE) at Nanyang Technological University (JHU IO 90071537), and the US National Science Foundation via CCRI Award \#2120435. We would like to thank the BLIP Lab transcription team for generating high quality ground-truth annotations, as well as Seraphina Fong and Le Tuan Anh for establishing effective data pipelines. 

\vfill\pagebreak
\bibliographystyle{IEEEtran}
\bibliography{mybib}

\begin{thebibliography}{10}
\providecommand{\url}[1]{#1}
\csname url@samestyle\endcsname
\providecommand{\newblock}{\relax}
\providecommand{\bibinfo}[2]{#2}
\providecommand{\BIBentrySTDinterwordspacing}{\spaceskip=0pt\relax}
\providecommand{\BIBentryALTinterwordstretchfactor}{4}
\providecommand{\BIBentryALTinterwordspacing}{\spaceskip=\fontdimen2\font plus
\BIBentryALTinterwordstretchfactor\fontdimen3\font minus
  \fontdimen4\font\relax}
\providecommand{\BIBforeignlanguage}[2]{{%
\expandafter\ifx\csname l@#1\endcsname\relax
\typeout{** WARNING: IEEEtran.bst: No hyphenation pattern has been}%
\typeout{** loaded for the language `#1'. Using the pattern for}%
\typeout{** the default language instead.}%
\else
\language=\csname l@#1\endcsname
\fi
#2}}
\providecommand{\BIBdecl}{\relax}
\BIBdecl

\bibitem{lre_eval}
S.~O. Sadjadi, T.~Kheyrkhah, C.~S. Greenberg, E.~Singer, D.~A. Reynolds, L.~P.
  Mason, and J.~Hernandez-Cordero, ``{P}erformance {A}nalysis of the 2017
  {NIST} {L}anguage {R}ecognition {E}valuation,'' in \emph{Proc. Interspeech},
  2018, pp. 1798--1802.

\bibitem{efficient_lid}
H.~Liu, L.~P.~G. Perera, A.~W.~H. Khong, E.~S. Chng, S.~J. Styles, and
  S.~Khudanpur, ``{E}fficient {S}elf-{S}upervised {L}earning {R}epresentations
  for {S}poken {L}anguage {I}dentification,'' \emph{IEEE J. Sel. Topics Signal
  Process.}, vol.~16, no.~6, pp. 1296--1307, 2022.

\bibitem{librispeech}
V.~Panayotov, G.~Chen, D.~Povey, and S.~Khudanpur, ``{L}ibrispeech: {A}n {ASR}
  {c}orpus {B}ased on {P}ublic {D}omain {A}udio {B}ooks,'' in \emph{Proc. IEEE
  Int. Conf. Acoust., Speech, Signal Process.}, 2015, pp. 5206--5210.

\bibitem{ami}
W.~Kraaij, T.~Hain, M.~Lincoln, and W.~Post, ``{T}he {AMI} {M}eeting
  {C}orpus,'' 2005.

\bibitem{chime5}
J.~Barker, S.~Watanabe, E.~Vincent, and J.~Trmal, ``{The Fifth 'CHiME' Speech
  Separation and Recognition Challenge: Dataset, Task and Baselines},'' in
  \emph{Proc. Interspeech 2018}, 2018, pp. 1561--1565.

\bibitem{kukk22_interspeech}
K.~Kukk and T.~Alumäe, ``{{I}mproving {L}anguage {I}dentification of
  {A}ccented {S}peech},'' in \emph{Proc. Interspeech}, 2022, pp. 1288--1292.

\bibitem{chan22b_interspeech}
M.~P.~Y. Chan, J.~Choe, A.~Li, Y.~Chen, X.~Gao, and N.~Holliday, ``{{T}raining
  and {T}ypological {B}ias in {ASR} {P}erformance for {W}orld {E}nglishes},''
  in \emph{Proc. Interspeech}, 2022, pp. 1273--1277.

\bibitem{zhang22n_interspeech}
Y.~Zhang, Y.~Zhang, B.~Halpern, T.~Patel, and O.~Scharenborg, ``{{M}itigating
  {B}ias {A}gainst {N}on-{N}ative {A}ccents},'' in \emph{Proc. Interspeech
  2022}, 2022, pp. 3168--3172.

\bibitem{liu21_interspeech}
H.~Liu, L.~P.~G. Perera, X.~Zhang, J.~Dauwels, A.~W.~H. Khong, S.~Khudanpur,
  and S.~J. Styles, ``{E}nd-to-{E}nd {L}anguage {D}iarization for {B}ilingual
  {C}ode-{S}witching {S}peech,'' in \emph{Proc. Interspeech}, 2021, pp.
  1489--1493.

\bibitem{liu22_interspeech}
H.~Liu, L.~P.~G. Perera, A.~W.~H. Khong, S.~J. Styles, and S.~Khudanpur,
  ``{PHO-LID}: {A} {U}nified {M}odel {I}ncorporating {A}coustic-{P}honetic and
  {P}honotactic {I}nformation for {L}anguage {I}dentification,'' in \emph{Proc.
  Interspeech}, 2022, pp. 2233--2237.

\bibitem{liu22c_odyssey}
H.~Liu, L.~P.~G. Perera, A.~W.~H. Khong, J.~Dauwels, S.~J. Styles, and
  S.~Khudanpur, ``{E}nhancing {L}anguage {I}dentification {U}sing {D}ual-{M}ode
  {M}odel with {K}nowledge {D}istillation,'' in \emph{Proc. Odyssey Speaker
  Lang. Recognit. Workshop}, 2022, pp. 248--254.

\bibitem{bi_encoder}
Y.~Lu, M.~Huang, H.~Li, J.~Guo, and Y.~Qian, ``{{B}i-encoder {T}ransformer
  {N}etwork for {M}andarin-{E}nglish {C}ode-switching {S}peech {R}ecognition
  {U}sing {M}ixture of {E}xperts},'' in \emph{Proc. Interspeech}, 2020, pp.
  4766--4770.

\bibitem{cristia18_interspeech}
A.~Cristia, S.~Ganesh, M.~Casillas, and S.~Ganapathy, ``{{T}alker {D}iarization
  in the {W}ild: the {C}ase of {C}hild-{C}entered {D}aylong
  {A}udio-{R}ecordings},'' in \emph{Proc. Interspeech 2018}, 2018, pp.
  2583--2587.

\bibitem{cristia2021thorough}
A.~Cristia, M.~Lavechin, C.~Scaff, M.~Soderstrom, C.~Rowland,
  O.~R{\"a}s{\"a}nen, J.~Bunce, and E.~Bergelson, ``{A} {T}horough {E}valuation
  of the {L}anguage {E}nvironment {A}nalysis (lena) {S}ystem,'' \emph{Behavior
  Research Methods}, vol.~53, no.~2, pp. 467--486, 2021.

\bibitem{UTHER20072}
M.~Uther, M.~Knoll, and D.~Burnham, ``{D}o {Y}ou {S}peak {E-N-G-L-I-S-H}? {A}
  {C}omparison of {F}oreigner- and {I}nfant-{D}irected {S}peech,'' \emph{Speech
  Communication}, vol.~49, no.~1, pp. 2--7, 2007.

\bibitem{robot}
S.~Perez-Kriz, G.~Anderson, and J.~Trafton, ``{R}obot-{D}irected {S}peech:
  {U}sing {L}anguage to {A}ssess {F}irst-time {U}sers' {C}onceptualizations of
  a {R}obot,'' 03 2010, pp. 267--274.

\bibitem{hazan2015}
V.~Hazan, M.~Uther, and S.~Granlund, ``{H}ow {D}oes {F}oreigner-{D}irected
  {S}peech {D}iffer {F}rom {O}ther {F}orms of {L}istener-{D}irected {C}lear
  {S}peaking {S}tyles?'' 08 2015.

\bibitem{Burnham2002}
D.~Burnham, C.~Kitamura, and U.~Vollmer-Conna, ``{W}hat's {N}ew, {P}ussycat?
  {O}n {T}alking to {B}abies and {A}nimals,'' \emph{Science (New York, N.Y.)},
  vol. 296, p. 1435, 06 2002.

\bibitem{merliondata}
Y.~H.~V. Chua, L.~P. Garcia~Perera, S.~Khudanpur, A.~W.~H. Khong, J.~Dauwels,
  F.~T. Woon, and S.~J. Styles, ``{D}evelopment and {E}valuation {d}ata for
  {M}ultilingual {E}veryday {R}ecordings - {L}anguage {I}dentification on
  {C}ode-{S}witched {C}hild-{D}irected {S}peech ({MERLI}on {CCS})
  {C}hallenge,'' 2023, {D}R-NTU (Data), V1
  \url{https://doi.org/10.21979/N9/ANXS8Z}.

\bibitem{woon2021creating}
F.~T. Woon, E.~C. Yogarrajah, S.~Fong, N.~S.~M. Salleh, S.~Sundaray, and S.~J.
  Styles, ``{C}reating a {C}orpus of {M}ultilingual {P}arent-child {S}peech
  {R}emotely: {L}essons {L}earned in a {L}arge-scale {O}nscreen {P}icturebook
  {S}haring {T}ask,'' \emph{Frontiers in Psychology}, vol.~12, 2021.

\bibitem{styles2020little}
S.~J. Styles, ``Little {O}rangutan: {W}hat a {S}cary {S}torm,'' 2020, {DR-NTU}
  (Data), V1, \url{https://doi.org/10.21979/N9/MJMFXV}.

\bibitem{bela}
A.~Loh, E.~C. Yogarrajah, F.~T. Woon, J.~Wong, N.~S. Mohd~Salleh, S.~Fong,
  S.~Sundaray, S.~Binte~Amran, S.~J. Styles, T.~A. Le, Y.~H.~V. Chua, and
  V.~Selvarajan, ``{BLIP} {L}ab {ELAN} {L}anguage {A}nnotation ({BELA})
  {T}ranscription {C}onventions.'' \url{https://blipntu.github.io/belacon/},
  2021.

\bibitem{deterding2007singapore}
D.~Deterding, \emph{Singapore English}.\hskip 1em plus 0.5em minus 0.4em\relax
  Edinburgh University Press, 2007.

\bibitem{lee2010tonal}
L.~Lee, ``{The {T}onal {S}ystem of {S}ingapore {M}andarin},'' in \emph{Proc. of
  NACCL}, 2010, pp. 345--362.

\bibitem{dihard}
N.~Ryant, K.~W. Church, C.~Cieri, A.~Cristia, J.~Du, S.~Ganapathy, and M.~Y.
  Liberman, ``{T}he {S}econd {DIHARD} {D}iarization {C}hallenge: {D}ataset,
  {T}ask, and {B}aselines,'' in \emph{Proc. Interspeech}, 2019.

\bibitem{wang22b_odyssey}
Q.~Wang, Y.~Yu, J.~Pelecanos, Y.~Huang, and I.~L. Moreno, ``{{A}ttentive
  {T}emporal {P}ooling for {C}onformer-{B}ased {S}treaming {L}anguage
  {I}dentification in {L}ong-{F}orm {S}peech},'' in \emph{Proc. The Speaker and
  Language Recognition Workshop (Odyssey 2022)}, 2022, pp. 255--262.

\bibitem{evalplan}
L.~P. Garcia~Perera, Y.~H.~V. Chua, H.~Liu, F.~T. Woon, A.~W.~H. Khong,
  J.~Dauwels, and S.~J. Styles, ``{MERLI}on {CCS} {C}hallenge {E}valuation
  {P}lan {V}ersion 1.2,'' February 17, 2023, {A}vailable:
  https://bit.ly/merlion-ccs-eval-plan-v1-2.

\bibitem{codalab_competitions}
\BIBentryALTinterwordspacing
A.~Pavao, I.~Guyon, A.-C. Letournel, X.~Baró, H.~Escalante, S.~Escalera,
  T.~Thomas, and Z.~Xu, ``Codalab competitions: An open source platform to
  organize scientific challenges,'' \emph{Technical report}, 2022. [Online].
  Available: \url{https://hal.inria.fr/hal-03629462v1}
\BIBentrySTDinterwordspacing

\bibitem{whisper}
\BIBentryALTinterwordspacing
A.~Radford, J.~W. Kim, T.~Xu, G.~Brockman, C.~McLeavey, and I.~Sutskever,
  ``{R}obust {S}peech {R}ecognition via {L}arge-{S}cale {W}eak {S}upervision,''
  2022. [Online]. Available: \url{https://arxiv.org/abs/2212.04356}
\BIBentrySTDinterwordspacing

\bibitem{wav2vec}
A.~Baevski, Y.~Zhou, A.~Mohamed, and M.~Auli, ``wav2vec 2.0: {A} {F}ramework
  for {S}elf-{S}upervised {L}earning of {S}peech {R}epresentations,''
  \emph{Proc. Adv. Neural Inf. Process. Syst.}, vol.~33, pp. 12\,449--12\,460,
  2020.

\bibitem{titanet}
N.~R. Koluguri, T.~Park, and B.~Ginsburg, ``{T}itanet: {N}eural {M}odel for
  {S}peaker {R}epresentation with 1{D} {D}epth-wise {S}eparable {C}onvolutions
  and {G}lobal {C}ontext,'' in \emph{Proc. IEEE Int. Conf. Acoust., Speech,
  Signal Process.}, 2022, pp. 8102--8106.

\bibitem{nemo}
\BIBentryALTinterwordspacing
O.~Kuchaiev, J.~Li, H.~Nguyen, O.~Hrinchuk, R.~Leary, B.~Ginsburg, S.~Kriman,
  S.~Beliaev, V.~Lavrukhin, J.~Cook, P.~Castonguay, M.~Popova, J.~Huang, and
  J.~M. Cohen, ``Nemo: {A} {T}oolkit for {B}uilding {AI} {A}pplications {U}sing
  {N}eural {M}odules,'' \emph{CoRR}, vol. abs/1909.09577, 2019. [Online].
  Available: \url{http://arxiv.org/abs/1909.09577}
\BIBentrySTDinterwordspacing

\bibitem{ardila-etal-2020-common}
R.~Ardila, M.~Branson, K.~Davis, M.~Kohler, J.~Meyer, M.~Henretty, R.~Morais,
  L.~Saunders, F.~Tyers, and G.~Weber, ``{C}ommon {V}oice: {A}
  {M}assively-{M}ultilingual {S}peech {C}orpus,'' in \emph{Proceedings of the
  Twelfth Language Resources and Evaluation Conference}.\hskip 1em plus 0.5em
  minus 0.4em\relax Marseille, France: European Language Resources Association,
  May 2020, pp. 4218--4222.

\bibitem{discoursemarkers}
D.~Smakman and S.~Wagenaar, ``{D}iscourse {P}articles in {C}olloquial
  {S}ingapore {E}nglish,'' \emph{World Englishes}, vol.~32, no.~3, pp.
  308--324, 2013.

\bibitem{merlionerrors}
S.~J. Styles, Y.~H.~V. Chua, F.~T. Woon, H.~Liu, L.~P. Garcia~Perera,
  S.~Khudanpur, A.~W.~H. Khong, and J.~Dauwels, ``Investigating model
  performance in language identification: beyond simple error statistics,'' in
  \emph{Proc. Interspeech}, 2023.

\end{thebibliography}

\end{document}